\pdfoutput=1
\documentclass[prl,aps,reprint,showpacs,showkeys,floats,floatfix,superscriptaddress, citeautoscript]{revtex4-1}
\usepackage{natbib}
\setcitestyle{super}
\AtBeginDocument{\nocite{achemso-control}}

\usepackage{graphicx}
\usepackage{dcolumn}
\usepackage[colorlinks,allcolors=black,citecolor=blue,urlcolor=blue]{hyperref}
\usepackage{textcomp, color}

\begin{document}

\title{Nuclear Quantum Effects in Liquid Water Are Negligible for Structure\\ but Significant for Dynamics}

\author{Nore Stolte}
\email{nore.stolte@theochem.rub.de\\janos.daru@ttk.elte.hu}
\affiliation{Lehrstuhl f{\"u}r Theoretische Chemie, Ruhr-Universit{\"a}t Bochum, 44780 Bochum, Germany}
\author{J{\'a}nos Daru}
\email{nore.stolte@theochem.rub.de\\janos.daru@ttk.elte.hu}
\affiliation{Lehrstuhl f{\"u}r Theoretische Chemie, Ruhr-Universit{\"a}t Bochum, 44780 Bochum, Germany}
\affiliation{Department of Organic Chemistry, E{\"o}tv{\"o}s Lor{\'a}nd University, 1117 Budapest, Hungary}
\author{Harald Forbert}
\affiliation{Center for Solvation Science ZEMOS, Ruhr-Universit{\"a}t Bochum, 44780 Bochum, Germany}
\author{J{\"o}rg Behler}
\affiliation{Lehrstuhl f{\"u}r Theoretische Chemie II, Ruhr-Universit{\"a}t Bochum, 44780 Bochum, Germany}
\affiliation{Research Center Chemical Sciences and Sustainability, Research Alliance Ruhr, 44780 Bochum, Germany}
\author{Dominik Marx}
\affiliation{Lehrstuhl f{\"u}r Theoretische Chemie, Ruhr-Universit{\"a}t Bochum, 44780 Bochum, Germany}

\date{\today}

\begin{abstract}
Isotopic substitution, which can be realized both in experiment 
and
computer simulations, is a direct approach to assess the role of 
nuclear quantum effects on the structure and dynamics of matter. 
Yet, the impact of nuclear quantum effects on the structure
of liquid water as 
probed 
in experiment 
by comparing normal to 
heavy water 
has remained 
controversial.
To settle this issue, 
we employ a highly accurate machine-learned high-dimensional neural network potential 
to perform converged coupled cluster-quality path integral simulations
of 
liquid H$_2$O versus D$_2$O at ambient conditions. 
We find substantial H/D~quantum
effects on the rotational and translational dynamics of 
water, in close agreement with the experimental benchmarks.
However, 
in stark contrast
to the role for dynamics,
H/D 
quantum effects 
turn out to be unexpectedly small, on the order of 1/1000~{\AA}, on both 
intramolecular and \mbox{H-bonding} structure of water.
The most probable structure of water remains nearly unaffected by 
nuclear quantum effects, 
but effects on fluctuations away from average are appreciable, rendering H$_2$O 
substantially more 
``liquid" than D$_2$O.

\end{abstract}

\maketitle

Nuclear quantum effects (NQEs) play an important role in the properties of water 
due to the small nuclear mass of hydrogen atoms
participating in  
intermolecular bonds. 
The unique attributes of water are largely a result of its H-bonding network, so the
influence of NQEs on water
are 
expected to be
crucial for its chemical and physical characteristics.
Indeed, experiments
have identified many differences between 
H$_2$O and heavy water, 
D$_2$O, such as melting temperature, diffusion coefficient, surface tension, and density \cite{Ceriotti2016Nuclear}.
The electronic structure of H$_2$O and D$_2$O is identical~--
within the Born-Oppenheimer approximation~--
so these differences have to be attributed to the different 
masses of H versus D, 
and the 
larger extent of 
quantum delocalization in H$_2$O versus D$_2$O.
Simulation studies of liquid H$_2$O and D$_2$O that incorporate NQEs have aimed to pinpoint the 
differences between the two liquids at the molecular level~\cite{Ceriotti2016Nuclear}, but the 
structures 
of liquid H$_2$O and D$_2$O, and the differences between these two liquids due to NQEs, 
have remained elusive until today.

Structural isotope effects in liquid water have been 
extensively 
studied using scattering experiments \cite{Tomberli2000Isotopic, Hart2005Temperature, Soper2008Quantum, Zeidler2011Oxygen, Kim2017Temperature, 
Kameda2018Neutron}.
However, extracting real-space structural information from diffraction studies is challenging
\cite{Wikfeldt2009Range, Brookes2015Family}, leading to strong controversies on quantifying H/D isotope effects 
on the structure of water \cite{Soper2012Comment, Zeidler2012Reply,
Soper2013Radial}.
Furthermore, data analysis may be impeded by experimental challenges such as inelasticity effects \cite{Soper2013Radial, Kameda2021Direct}
resulting
in
different conclusions depending on data treatment. 
Consequently, experimental predictions for the difference in covalent bond 
lengths
between
H$_2$O and D$_2$O vary 
by orders of magnitude~\cite{Soper2008Quantum, Zeidler2011Oxygen, Kameda2018Neutron}, 
from \mbox{0.000 $\pm$ 0.001} to 0.03~{\AA}.
Even more importantly, for intermolecular bonds, 
pioneering 
isotope substitution experiments 
found a \emph{shorter} H-bonding distance in H$_2$O than D$_2$O \cite{Soper2008Quantum},
although in 
an improved 
analysis of all available data 
at that time \cite{Soper2013Radial}, 
the H-bonds in normal water were found to be \emph{longer} than in heavy water.
Thus, even
though experiments have consistently 
found that liquid D$_2$O is more structured than liquid H$_2$O \cite{Ceriotti2016Nuclear}, the
structural differences between 
liquid
H$_2$O and D$_2$O 
due to nuclear quantum effects remain a subject of vivid
discussion
up to the present time.
This concerns the order of magnitude of isotope effects on structure,
and even their sign.

In 
stark
contrast to structural properties, the
measured diffusion
coefficients~$D$ 
are unambiguous \cite{Weingartner1982Self}, with a ratio $D_{\mathrm{H_2O}} / D_{\mathrm{D_2O}} = 1.228 \pm 0.003$ \cite{Mills1973Self-diffusion}.
If differences between H$_2$O and D$_2$O were entirely 
classical 
due to trivial mass effects, then the diffusion coefficients would be related by $D_{\mathrm{H_2O}} = D_{\mathrm{D_2O}} \sqrt{m_{\mathrm{D_2O}} / m_{\mathrm{H_2O}}}$
\cite{Mills1976Effect}.
However, 
that mass ratio is 
very small, 
$\sqrt{m_{\mathrm{D_2O}} / m_{\mathrm{H_2O}}}\approx 1.05$
whereas the experimental diffusion coefficients differ by more than 20~\%, suggesting that differences in self-diffusion between H$_2$O and D$_2$O are governed not by mass 
but by NQEs.
Similarly, differences in reorientation dynamics of H$_2$O and D$_2$O are dominated by NQEs \cite{Wilkins2017Nuclear}, and 
all
experiments consistently find
$\tau_2(\mathrm{H_2O})/\tau_2(\mathrm{D_2O})$
\mbox{$\approx$ 0.8 $\pm$ 0.05} 
\cite{Jonas1976Molecular, Lankhorst1982Determination, Cringus2004Hydrogen, Rezus2005Orientational, Bakker2008Molecular, Qvist2012Rotational}. 
Thus, even though NQEs on the structure of liquid water,
as experimentally
accessible
by H/D isotope substitution, 
are 
controversial, they are well established for dynamical properties.

Computational 
investigations
of NQEs in liquid water 
have attracted significant attention and have
been ongoing since the 1980s, with 
early works that 
compared H$_2$O and D$_2$O \cite{Kuharski1985Quantum}, 
as well as
classical and quantum H$_2$O~\cite{Kuharski1984Quantum, Wallqvist1985Path-integral}.
Subsequently, force field and \textit{ab initio} molecular dynamics simulations have been used to study NQEs in liquid water
by directly comparing H$_2$O to D$_2$O in simulations that include NQEs
\cite{
DelBuono1991Model,
Chen2003Hydrogen, 
Paesani2009Properties,
Habershon2009Competing,
Paesani2011Hydrogen, Nagata2012Nuclear, Wilkins2017Nuclear, 
Machida2018Nuclear,
Ko2019Isotope,
Xu2020Isotope, Thomsen2022Structures},
or by comparing quantum H$_2$O to classical water \cite{Gai1996Classical, Morrone2008Nuclear, Paesani2010Nuclear, Kong2012Roles, Ceriotti2013Nuclear, Medders2014Development, Giberti2014Role, Zhang2021Modeling, Li2022Static, Yu2022QAQUA, Chen2023Data-Efficient, Yu2023Status}.
A discussion of 
the previous work
can be found in several reviews dedicated to this topic
\cite{Paesani2009Properties, Ceriotti2016Nuclear}.

Regardless of the
decades-long line of research, 
predictions of NQEs vary significantly across force field 
as well as 
density functional studies. 
For example, in one study, the first O--H peak in the radial distribution function (RDF) was found at smaller radial distance than the first O--D peak \cite{Xu2020Isotope}.
Similarly, for the qSPC/Fw water model \cite{Paesani2006Accurate} the O--H bond is shorter than the O--D bond \cite{Zeidler2012Isotope}, while the TTM3-F model \cite{Fanourgakis2008Development} predicted that the O--H bond is 0.5~\% longer than the O--D bond \cite{Zeidler2012Isotope}.
A study based on the PBE0-TS density functional even reported a 1~\% increase in the O--H bond length relative to O--D \cite{Ko2019Isotope}. 
Overall, computational studies disagree on the magnitude
and 
the sign of H/D isotope effects on the structure 
of liquid water~-- 
a situation akin to experiment. 

Despite the unclear experimental and computational situation as to
H/D isotope effects, the
differences between H$_2$O and D$_2$O have previously been qualitatively rationalized by the notion of competing NQEs \cite{Habershon2009Competing}
as follows:
NQEs enhance the delocalization of 
hydrogen atoms in H$_2$O compared to 
deuterium atoms in D$_2$O.
Delocalization
of H (D) atoms
along the axis connecting neighboring O~atoms weakens covalent O--H (O--D) bonds 
while it \emph{strengthens} intermolecular O$\cdots$H  (O$\cdots$D) bonds.
However, H (D) 
delocalization out of the axis connecting neighboring O atoms 
\emph{weakens} and eventually disrupts H-bonds \cite{Ceriotti2016Nuclear}.
Qualitatively,
there are two competing quantum effects on H-bond structure 
at work due to quantum delocalization of 
H (D) atoms \cite{Li2011Quantum}. 
However, the magnitude of the overall changes in structure due to competing NQEs is 
utmost
sensitive to the 
subtle balance of intermolecular versus intramolecular interactions in the particular force field or density functional used. Moreover, whether or not anharmonic effects in covalent bonds are described correctly plays an important role in determining that balance \cite{Habershon2009Competing, Ceriotti2016Nuclear}.
The 
differences of magnitude and sign in the reported isotope
effects 
on the structure of liquid water 
as obtained
from simulations
is likely due to different descriptions of these
delicate effects by different force fields and density functionals.
Evidently, such differences impact the quantitative 
predictions according to the concept of competing NQEs. 
Given the current situation, 
and despite the enormous body of work that exists,
it can be concluded that 
the impact of quantum effects on the structure
of water as probed by H/D isotope substitution is 
experimentally controversial and computationally 
unclear. 

\begin{figure}[b]
\includegraphics{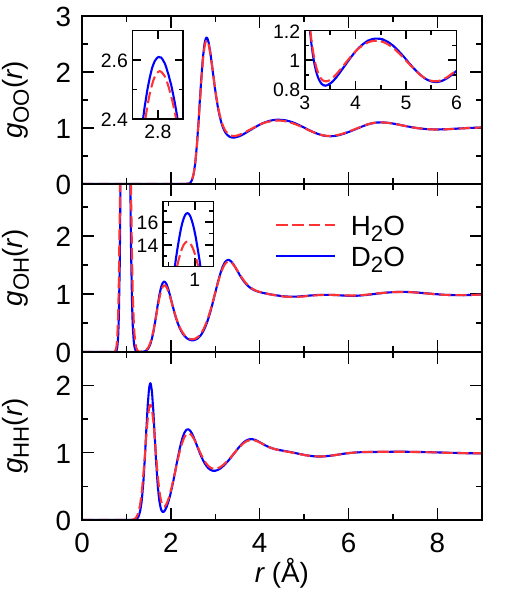}
\caption{Radial distribution functions of
normal
and heavy water from path integral CCMD simulations at 298 K, 
\mbox{1 atm.} 
The radial bin size is 0.005 {\AA} and distributions are not smoothened.}
\label{rdfs}
\end{figure}

Our recently developed approach \cite{Daru2022Coupled}, termed Coupled Cluster Molecular Dynamics (CCMD
for short),
uses a high-dimensional neural network potential (HDNNP)~\cite{Behler2007Generalized, Behler2021Four} to reproduce 
highly accurate CCSD(T) electronic structure theory \cite{Bartlett2007Coupled} in condensed phase simulations, 
at the computational cost typically associated with (advanced) force fields.
The perturbative triples correction in coupled cluster calculations,
(T),
is 
sufficient in order to achieve quantitative agreement with experiment for liquid water
\cite{Chen2023Data-Efficient}.
Our CCMD HDNNP was constructed with the aim of describing with high accuracy a 
specific
region in configuration space for water, namely that of bulk liquid water at ambient conditions
in the spirit of previous work on 
aqueous systems \cite{Schran2021Machine},
rather than generating a ``universal water potential''. 
For that reason, 
this
HDNNP was trained exclusively on configurations of bulk liquid water at ambient conditions, and 
its training set could potentially be extended when aiming to
study e.g.,  protonated water clusters, which were studied with a HDNNP constructed using distinct reference data \cite{KondatiNatarajan2015Representing, Schran2020Automated}.
In line with other coupled cluster studies 
of
liquid water 
\cite{Yu2022QAQUA, Chen2023Data-Efficient, Zhu2023MBpol, Yu2023Status}, 
excellent agreement has been found between the most reliable experimental 
H$_2$O
data available and CCMD \cite{Daru2022Coupled}.
With CCMD, it is both possible to simulate
water at CCSD(T) accuracy 
and to perform path integral 
simulations in the Trotter convergence limit that are long enough to 
finally 
settle the 
differences between H$_2$O and D$_2$O. 
Here, we use converged CCMD simulations relying on a grand total of 
more than 
60~ns 
of ring polymer molecular dynamics 
(RPMD)
simulations 
of 
H$_2$O and D$_2$O 
to determine isotope substitution effects in liquid water in an effort to conclusively resolve NQEs where previous studies 
had to remain ambiguous due to the high computational costs.

The RDFs of normal
and heavy water 
(Figure~\ref{rdfs}) 
turn out to be only slightly different.
In fact, the differences in H$_2$O and D$_2$O O--O RDFs are smaller than the spread of RDF predictions from an X-ray diffraction experiment for water~\cite{Brookes2015Family}.
Overall, the
O--H and H--H RDFs show 
the
expected softening of correlations in H$_2$O compared to D$_2$O, as a result of the enhanced quantum delocalization of 
protons ($^1$H$^+$) versus deuterons 
(D$^+$):
H$_2$O has lower peaks and shallower troughs than D$_2$O.

Firstly, we 
analyze the NQEs on
the intramolecular structure of H$_2$O versus D$_2$O
molecules. 
Using
a combination of
X-ray
and neutron scattering data \cite{Soper2008Quantum}, the O--H and O--D intramolecular bond lengths were found
to be 1.01 and 0.98 {\AA}, respectively,
providing 
a difference of 0.03~{\AA} in the 
average 
covalent bond length.
Based on reanalysis of all available data, it was later suggested that this 
difference is an overestimate \cite{Soper2013Radial}.
In oxygen isotope substitution neutron scattering experiments
\cite{Zeidler2011Oxygen},
peaks due to intramolecular correlations in the radial first difference functions were found at \mbox{0.990 $\pm$ 0.005 {\AA}} for H$_2$O, and at 
\mbox{0.985 $\pm$ 0.005 {\AA}} 
for D$_2$O, 
yielding
\mbox{$r_{\mathrm{OH}}-r_{\mathrm{OD}} = 0.005 \pm 0.007$ {\AA}.}
Yet, another
more recent neutron scattering investigation found no difference between the O--H and O--D covalent bond lengths \cite{Kameda2018Neutron}, putting both at \mbox{0.972 $\pm$ 0.001 {\AA},} 
though 
the O--D bond length was adjusted 
to \mbox{0.9759 $\pm$ 0.0007 {\AA}} in a later study \cite{Kameda2021Direct}, as inelasticity effects in the previous study led to an underestimated bond length.
From our CCMD simulations, the 
first peak in the O--H and O--D RDFs occur at 
0.9718(1) and 0.9717(1)~{\AA},
respectively, so the difference is 0.0001(1)~{\AA}.
The average intramolecular O--H and O--D distances, 
found by integrating over the first peak in the O--H and O--D distance distributions, 
are \mbox{$\left\langle d_{\mathrm{OH}}\right\rangle = 0.9888$ {\AA}} and \mbox{$\left\langle d_{\mathrm{OD}}\right\rangle = 0.9839$ {\AA},}
with errors beyond the reported precision,
thus \mbox{$\left\langle d_{\mathrm{OH}}\right\rangle - \left\langle d_{\mathrm{OD}}\right\rangle$ = 0.0049~{\AA}} or 4.9$\times$10$^{-3}$~{\AA}. 
Given this order of magnitude, we
conclude that the covalent bond length difference between H$_2$O and D$_2$O is 
on the order of 1/1000~{\AA},
thus supporting vanishingly small isotope effects on the covalent bond length 
in substantial accord with
the more recent experimental 
analyses.
Additionally, the average 
intramolecular 
bond angle remains 
virtually
unchanged upon 
H/D 
isotope substitution 
(see Supporting Information).

Secondly, we address the impact of NQEs on the intermolecular 
structure.
The differences in the oxygen correlations from RDFs 
depend only indirectly on 
proton
delocalization
as such, 
which is most significantly affected by NQEs.
Both the nearest-
and second-neighbor O--O RDF peaks are marginally higher for D$_2$O than for H$_2$O, and
the first peak is located at slightly smaller $r$ in D$_2$O.
Accordingly, the second O--H peak, at the typical H-bonding distance, is shifted to smaller $r$ 
in D$_2$O, 
where it 
is located at 1.8512(3)~{\AA}, versus 1.8572(2)~{\AA} in H$_2$O.
We thus find that the H-bonds in normal water are 
only
0.0060(4)~{\AA}
or 6.0$\times$10$^{-3}$~{\AA},
i.e., on the order of 1/1000~{\AA}
longer
compared to heavy water as a result of NQEs. 
In 
H/D~isotope substitution
experiments \cite{Soper2008Quantum}, the H$_2$O H-bonding peak was 
initially 
found at 
\mbox{$\approx$ 1.74 {\AA},} 
while it was shifted to a larger distance of 
\mbox{$\approx$ 1.81 {\AA}} in D$_2$O,
implying that H-bonds would be shorter by 0.07~{\AA} in normal versus heavy water. 
However, 
careful 
re-analysis based on all available data~\cite{Soper2013Radial} placed the H$_2$O intermolecular 
O$\cdots$H 
peak at the larger distance of 
\mbox{$\approx$ 1.82 {\AA},} while the O--D peak shifted to \mbox{$\approx$ 1.80 {\AA},}
implying that H-bonds in liquid H$_2$O are 
0.02~{\AA}
longer than in D$_2$O
(according to Figure~23 in Ref. \cite{Soper2013Radial}).
These most recent scattering results are thus more consistent with our findings for the H-bonding structure of H$_2$O 
versus
D$_2$O.
Using
oxygen substitution neutron scattering data \cite{Zeidler2011Oxygen}, the H-bonding peak was found at \mbox{1.83 $\pm$ 0.02 {\AA}} in D$_2$O, which is 
in agreement with our finding within experimental uncertainty
(while
the H$_2$O peak could not be resolved in the experiment).

\begin{figure}[b]
\includegraphics{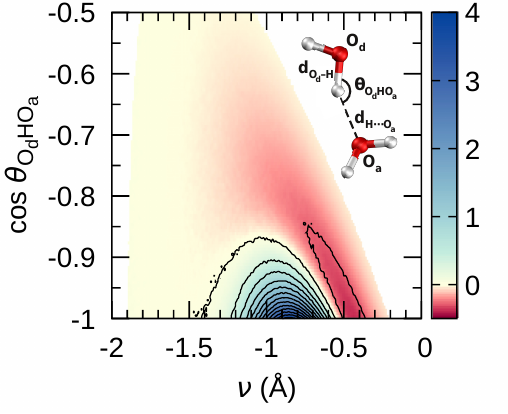}
\caption{
Difference of the normalized joint probability densities $p$ of H-bonding coordinates $\nu$ and $\cos  \theta_{\mathrm{O_dHO_a}} $ in D$_2$O 
versus
H$_2$O, $p({\mathrm{D_2O}})-p({\mathrm{H_2O}})$; 
$\nu = d_{\mathrm{O_d-H}}-d_{\mathrm{H\cdots O_a}}$ as defined in the inset together with $\theta_{\mathrm{O_dHO_a}}$.
O$_{\mathrm{d}}$ is the H-bond donor, O$_{\mathrm{a}}$ is the H-bond acceptor.
The plot is colored light yellow where 
$p({\mathrm{D_2O}})-p({\mathrm{H_2O}}) \approx 0$,
blue where the difference is positive, red where the difference is negative, 
and white where both individual probability densities are 
negligibly small,
$p <0.1$;
contour lines are drawn at intervals of~0.3.
The same figure with a color scale where the lightness gradient is the same for positive and negative values is presented in the 
Supporting Information.
}
\label{hb}
\end{figure}

We conclude that 
also the NQEs on the 
intermolecular O$\cdots$H distance, being the hallmark of H-bonding in liquid water,
are negligible, since they are only on the order of 1/1000~{\AA}. 
Similarly, the
tetrahedral order parameter and oxygen triplet angle distributions 
also reveal that local orientational ordering is insensitive to NQEs, D$_2$O 
being 
only sligthly more structured than H$_2$O 
(see Supporting Information).
Overall, we find 
only 
marginal 
differences between the intermolecular structure of 
liquid
H$_2$O versus D$_2$O.

Given the accuracy of CCMD and the 
high 
statistical convergence of our simulations, 
we can now quantify and thus scrutinize the impact of 
competing NQEs on the structure of 
liquid H$_2$O versus D$_2$O.
The joint probability density of the H-bonding angle and proton transfer coordinate \cite{Wang2014Quantum}
shows the distribution of protons 
or deuterons
in H-bonds in our simulations 
(see Supporting Information),
thus their difference
quantifies
NQEs on H-bonding structure (Figure~\ref{hb}).
Where the difference is negative
(red area), 
the probability density of H$_2$O is larger than that of D$_2$O
as a result of enhanced NQEs. 
Accordingly, H-bonds with less negative $\nu$ and larger
$\cos \theta_{\mathrm{O_dHO_a}}$
are more likely in H$_2$O than in D$_2$O,
i.e., in H$_2$O protons are more delocalized along the H-bonding axis and H-bond bending is 
also
enhanced.
The former strengthens H-bonds, while the latter weakens H-bonds \cite{Habershon2009Competing}.
Our evidence suggests
that the magnitude of these two 
nuclear
delocalization effects 
is \emph{nearly} equal
in liquid water,  
to render 
``the more quantum-mechanical water'', namely 
liquid H$_2$O, structurally very similar to heavy water, with D$_2$O being 
only
\emph{slightly} more structured than H$_2$O,
thus transcending previous analyses.

\begin{figure}[b]
\includegraphics{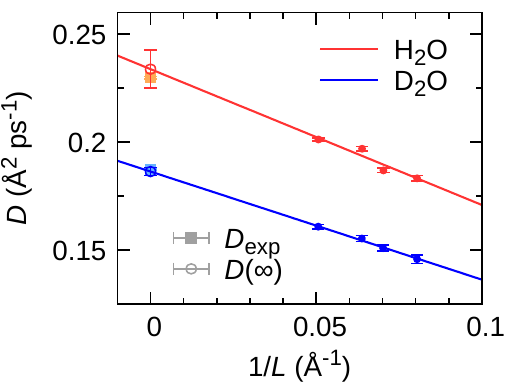}
\caption{Self-diffusion coefficient $D$ of H$_2$O and D$_2$O at 298~K extrapolated 
to infinite system size using four different supercell sizes,
where $L$ is the cubic box 
length.
The experimental diffusion coefficients \cite{Mills1973Self-diffusion}
are shown with pale red and blue squares.
}
\label{diffusion}
\end{figure}

Translational diffusion in liquid water is highly dependent on the H-bonding environment and its dynamics, so the comparison of self-diffusion coefficients $D$ of H$_2$O and D$_2$O provides an excellent measure of the quantum nature of the H-bonding network.
Furthermore, 
isotope effects on $D$
provide 
arguably
the most uncontroversial experimental quantification of NQEs in liquid water,
as opposed to
experimental results for structure which 
remain
elusive.
We therefore consider $D$ and its H/D isotope effect as 
stringent 
benchmarks
to assess the quality of our CCMD simulations including the NQEs. 
In simulations, $D$ is subject to finite-size effects and 
is obtained here from explicit system-size extrapolation (Figure~\ref{diffusion}; 
see Supporting Information).
We find $D$ of H$_2$O and D$_2$O at 298~K to be
\mbox{0.234 $\pm$ 0.009} 
and \mbox{0.186 $\pm$ 0.002 
{\AA}$^2$ ps$^{-1}$}, 
respectively, in excellent agreement with the experimental results of \mbox{0.2299 $\pm$ 0.0005 {\AA}$^2$ ps$^{-1}$} for H$_2$O and \mbox{0.1872 $\pm$ 0.0004 {\AA}$^2$ ps$^{-1}$} for D$_2$O~\cite{Mills1973Self-diffusion}.
This benchmarking 
validates CCMD
and thereby supports strongly our conclusions above about NQEs on structural correlations 
and H-bonding
in liquid water.  
We note in passing that reproducing the experimentally well-established diffusion coefficients strongly supports the general consensus that RPMD provides reliable quantum effects on dynamical properties in liquid water at ambient conditions~\cite{Craig2004Quantum, Miller2005Quantum, Habershon2013Ring, Ceriotti2016Nuclear}.

\begin{table}[b]
\centering
\caption{Orientational relaxation times (ps)
of the O--H bond vector, $\tau_2^{\mathrm{OH}}$. CCMD results were obtained at 298 K. Experimental results labeled ``NMR" are obtained from NMR relaxation experiments, while 
``IR" refers to IR pump-probe experiments
(see Supporting Information).}
\label{reorientation}
\resizebox{8.6cm}{!}{%
\setlength\extrarowheight{2.5pt}
\begin{tabular}{l c c c} 
\hline
\hline
 & H$_2$O & D$_2$O & ${\tau_2^{\mathrm{OH}}}/{\tau_2^{\mathrm{OD}}}$ \\
\hline
CCMD integral & 1.796 $\pm$ 0.013 & 2.353 $\pm$ 0.031 & 0.763 $\pm$ 0.012 \\
NMR \cite{Jonas1976Molecular} & $\approx$ 2.5 & $\approx$ 3.0 & $\approx$ 0.8 \\
NMR \cite{Lankhorst1982Determination} & 1.71 $\pm$ 0.07 & 2.19 $\pm$ 0.10 & 0.78 $\pm$ 0.01 \\
NMR \cite{Qvist2012Rotational} & 1.83 $\pm$ 0.05 & 2.21 $\pm$ 0.06 & 0.83 $\pm$ 0.03 \\
\hline
CCMD fit & 3.046 $\pm$ 0.047 & 3.798 $\pm$ 0.053 & 0.802 $\pm$ 0.017 \\
IR \cite{Cringus2004Hydrogen} & & 3.0 $\pm$ 0.5 & \\
IR \cite{Rezus2005Orientational} & 2.5 $\pm$ 0.1 &  &  \\
IR \cite{Bakker2008Molecular} & 2.5 $\pm$ 0.2 & 3.0 $\pm$ 0.3 & 0.8 $\pm$ 0.1 \\
\hline
\hline 
\end{tabular}%
}
\end{table}

From 
CCMD, we find 
\mbox{$D_{\mathrm{H_2O}} / D_{\mathrm{D_2O}} = 1.25 \pm 0.05$} 
which compares favorably 
to 
experiment: 
\mbox{$1.228 \pm 0.003$} \cite{Mills1973Self-diffusion}.
Thus, diffusion of liquid H$_2$O relative to D$_2$O is very much faster 
than expected based on trivial mass effects,
suggesting that NQEs are of 
significant 
importance to the translational dynamics in liquid water.

The orientational relaxation of water molecules is 
intimately connected to H-bond breaking and making dynamics
\cite{Laage2006Molecular}.
Moreover, it has been demonstrated that NQEs are crucial to 
correctly describe the rotational dynamics~\cite{Wilkins2017Nuclear}.
The orientational relaxation times $\tau_2$ of the O--H bond vector 
from CCMD are compiled in Table~\ref{reorientation} and compared to 
IR and NMR reference data 
(see Supporting Information).
While different experiments give different 
$\tau_2$ values,
their O--H versus O--D \emph{ratio},
and thus the dynamical NQE on water rotational motion, 
is consistently similar across all studies, 
around 0.8, and thus in 
agreement with our CCMD data.
Additionally, 
that
$\tau_2^{\mathrm{OH}}/\tau_2^{\mathrm{OD}}$ ratio is very similar to the 
inverse diffusion coefficient ratio,
\mbox{0.80},
which indicates that NQEs affect translational and rotational motion similarly  \cite{Wilkins2017Nuclear}. 
We conclude that we find substantial 
NQEs on H-bond dynamics.

\begin{figure}[b]
\includegraphics{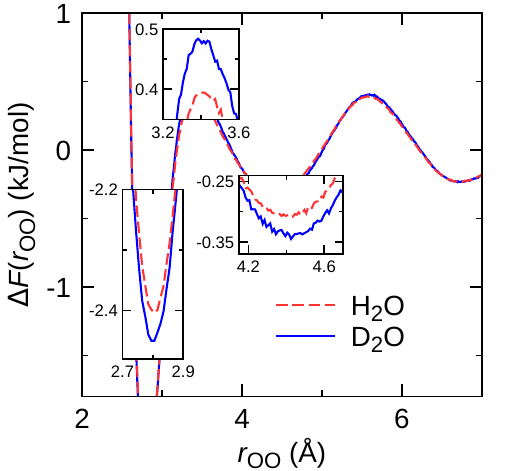}
\caption{Free energy profile along the O--O intermolecular distance.
The radial bin size is 0.01 {\AA} and the curves are not smoothened. 
}
\label{pmf}
\end{figure}

We have shown that  H$_2$O and D$_2$O are nearly identical structurally, indicating that positions of free energy minima are nearly unchanged due to NQEs, while the dynamics of H$_2$O are significantly enhanced compared to D$_2$O beyond what is expected from classical mass effects.
Dynamics are sensitive to free energy barriers, rather than to free energy minima positions
which determine the most probable structure.
We find significant differences between H$_2$O and D$_2$O in the free energy barrier heights along the intermolecular \mbox{O--O} distance (Fig.~\ref{pmf}, Table~\ref{barriers}).
How can these differences in activation free energies of normal versus heavy water be connected to dynamics?
First of all, 
H-bond exchange 
has been shown to be 
closely tied to water self-diffusion~\cite{Gomez2022Water}.
Secondly, 
the free energy barrier to H-bond exchange 
has been shown to be approximately given by 
the sum of free energy barriers associated with elongation of one \mbox{O--O} distance, and contraction of another \mbox{O--O} distance~\cite{Wilkins2017Nuclear},
which corresponds to the initial H-bond acceptor moving away from the H-bond donor and a new H-bond acceptor moving towards the H-bond donor.
Using that approximation of the energetic barrier to H-bond exchange, we estimate based on approximate quantum transition state theory that the ratio of H-bond exchange rates between H$_2$O and D$_2$O is \mbox{1.108 $\pm$ 0.002}
due to 
the
differences in intermolecular free energy barriers
reported in Table~\ref{barriers}.
The more pronounced fluctuations of the light protons in H$_2$O ``soften" the free energy landscape, lowering free energy barriers and decreasing activation free energies, and thus significantly 
enhance dynamics in H$_2$O compared to D$_2$O
while free energy minima positions are largely unaffected 
consistent with structural isotope effects of about one permille.

\begin{table}[t]
\centering
\caption{Activation free energies $\Delta F^{\ddagger}$ in the intermolecular O--O coordinate, in kJ/mol.
$\Delta F^{\ddagger}$ is the free energy difference between the bottom of the well and the top of the barrier.
``Min.~1 $\rightarrow$ min.~2" refers to $\Delta F^{\ddagger}$ when passing from the first to the second minimum in the free energy profile (Fig.~\ref{pmf}), and ``min.~2 $\rightarrow$ min.~1" is $\Delta F^{\ddagger}$ when going over the barrier in the opposite direction. 
}
\label{barriers}
\setlength\extrarowheight{2.5pt}
\begin{tabular}{l c c} 
\hline
\hline
 & $\Delta F^{\ddagger}_{\mathrm{H_2O}}$ & $\Delta F^{\ddagger}_{\mathrm{D_2O}}$  \\
 \hline 
Min. 1 $\rightarrow$ min. 2 &$\;\;$2.798 $\pm$ 0.002$\;\;$&$\;\;$2.932 $\pm$ 0.003$\;\;$\\
Min. 2 $\rightarrow$ min. 1 & 0.701 $\pm$ 0.001 & 0.821 $\pm$ 0.002 \\
\hline
\end{tabular}%
\end{table}

Based on 
converged path integral 
simulations of bulk liquid H$_2$O and D$_2$O at coupled cluster 
accuracy, CCMD, 
we find compelling agreement between the computed 
significant NQEs on \emph{dynamics} and the experimental 
benchmark
data.
This validation provides strong support for the reliability of our
hitherto unknown 
quantitative insights regarding NQEs on the \emph{structure} of liquid water: 
They are marginal with only negligible differences 
between liquid H$_2$O and D$_2$O 
on the order of 1/1000~{\AA}. 
Remarkably, 
even
the H-bond length in H$_2$O is only on the order of 0.1~{\%}
longer than in D$_2$O, 
making heavy water only a
marginally
more structured liquid. 
These 
findings disclose that competing NQEs
render
normal and heavy water structurally extremely similar liquids,
yet their dynamics is significantly different 
with reference to classical mass-ratio 
effects, 
as a result of non-trivial quantum effects
on free energy barriers.
Enhanced quantum fluctuations in H$_2$O lead to significant effects on free energy barriers, rendering H$_2$O more liquid than D$_2$O. 
This is traced back to the fluctuations away from the average, or most probable, structures.
We see a $\sim$~10~\% increase in H-bond exchange rates in H$_2$O relative to D$_2$O based on free energy barrier heights, compared to the 
permille difference $\sim$~0.1~\% 
on structure.
We anticipate that these challenging findings will stimulate 
novel experiments to provide much more reliable structural data
not just for neat water as an important reference liquid, but for 
aqueous solutions in general. 

\section{Methods}
We performed path integral simulations \cite{MarxHutter2009} of H$_2$O and D$_2$O 
at 298~K and 1~atm 
with 32~Trotter replica
using the CP2k software \cite{CP2k} with its path integral module \cite{Brieuc2020Converged} in conjunction with the HDNNP module \cite{Schran2018High}.
From the simulations, we compute 
those
structural and dynamical properties that can be extracted rather accurately from experiment, providing meaningful comparisons.
Structural properties and orientational relaxation times 
were extracted from extensive RPMD simulations with 256 molecules,  
whereas 
64, 96, 128, and 256 molecules were used to extrapolate the diffusion coefficients to 
infinite box size (see Supporting Information).

\section{Acknowledgments}
We are most grateful to Alan Soper and Yasuo Kameda for helpful discussions
and in particular to Alan Soper for having provided the numerical data 
underlying Figure~23 in Ref. 12. 
Funded by the Deutsche Forschungsgemeinschaft
(DFG, German Research Foundation) under Germany's
Excellence Strategy~-- EXC~2033~-- 390677874~-- RESOLV.
Financial support from the National Research, Development
and Innovation Office (NKFIH, Grant No. FK147031) is gratefully acknowledged
by JD.
All computations have been carried out locally at
HPC@ZEMOS, HPC-RESOLV, and BOVILAB@RUB.

\setlength{\bibsep}{0.0cm}
\bibliography{isotope_effects}

\end{document}


\title{Supporting Information for:\\
\vspace{0.4 cm}
\centerline{
\mbox{Nuclear Quantum Effects in Liquid Water Are Negligible for Structure}} but Significant for Dynamics}

\author{Nore Stolte}
\affiliation{Lehrstuhl f{\"u}r Theoretische Chemie, Ruhr-Universit{\"a}t Bochum, 44780 Bochum, Germany}
\author{J{\'a}nos Daru}
\affiliation{Lehrstuhl f{\"u}r Theoretische Chemie, Ruhr-Universit{\"a}t Bochum, 44780 Bochum, Germany}
\affiliation{Department of Organic Chemistry, E{\"o}tv{\"o}s Lor{\'a}nd University, 1117 Budapest, Hungary}
\author{Harald Forbert}
\affiliation{Center for Solvation Science ZEMOS, Ruhr-Universit{\"a}t Bochum, 44780 Bochum, Germany}
\author{J{\"o}rg Behler}
\affiliation{Lehrstuhl f{\"u}r Theoretische Chemie II, Ruhr-Universit{\"a}t Bochum, 44780 Bochum, Germany}
\affiliation{Research Center Chemical Sciences and Sustainability, Research Alliance Ruhr, 44780 Bochum, Germany}
\author{Dominik Marx}
\affiliation{Lehrstuhl f{\"u}r Theoretische Chemie, Ruhr-Universit{\"a}t Bochum, 44780 Bochum, Germany}

\maketitle

\tableofcontents

\clearpage

\section{Path Integral Molecular Dynamics Simulations}
Our simulations 
of liquid water 
at 298~K and 1~atm 
were performed
at constant volume with the experimental density \cite{Wagner2000IAPWS, Herrig2018Reference}
using CP2k \cite{CP2k, Hutter2014CP2K}.
To this end, we used the path integral module \cite{Brieuc2020Converged} 
of CP2k
together with the HDNNP module \cite{Schran2018High} which is compatible with the output file format of RubNNet4MD \cite{RubNNet4MD}, which was used to train the CCMD HDNNP \cite{Daru2022Coupled};
the 
MM radii of O and H were set to 0.380 and 0.450~{\AA}, respectively, which were communicated to CP2k via the MM\_RADIUS keyword.
The path integral molecular dynamics simulations \cite{MarxHutter2009} were performed with 32~Trotter replica and a time step of 0.25~fs.
In simulations, the H mass was 1.00798175 amu and the D mass was 2.01410178 amu
while the 
O mass was 15.9993047 amu.
Results for structure and orientational relaxation times presented in the main text were obtained from ring polymer molecular dynamics (RPMD) simulations 
\cite{Craig2004Quantum,Habershon2013Ring} with 256 water molecules.
The self-diffusion coefficients were obtained from RPMD simulations with 64, 96, 128, and 256 molecules 
to allow for finite-size scaling as explained below.
For all RPMD simulations, initial 
phase space
conditions were drawn from thermostatted path integral simulations, with the path integral Langevin equation (PILE) thermostatting approach \cite{Ceriotti2010Efficient}. 
For D$_2$O, 5~independent 100~ps long PILE simulations were performed for each system size.
From these, 50~initial conditions for RPMD simulations were sampled. 
Subsequently, 100~ps long RPMD simulations were performed for each initial condition, adding up to 5~ns of total RPMD simulation time for each of the system sizes.
For H$_2$O, a total of 
11~independent 100~ps long PILE simulations were performed for each system size, giving 110~initial configurations for a total of 11~ns of RPMD simulations per system size.
Thus, all
properties have been computed from the RPMD simulations
to provide the same data base for the structural and dynamical
observables.

\clearpage
\section{Intramolecular Bond Angle}
\begin{figure}[h!]
\includegraphics{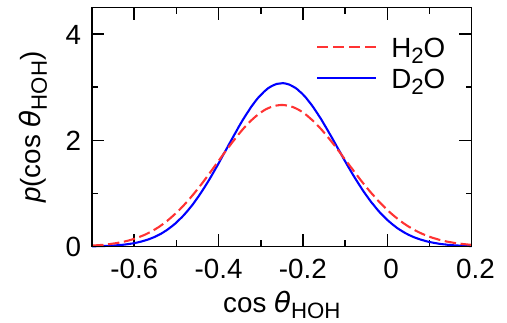}
\caption{Probability density of the cosine of the intramolecular bond angle of H$_2$O and D$_2$O.}
\label{angle}
\end{figure}

The probability distribution function of the intramolecular bond angle is broader in H$_2$O than in D$_2$O, as might be expected on account of the 
more pronounced 
quantum nature of H versus D~nuclei (Figure~\ref{angle}). 
Yet, the average bond angle is 
found to be
indistinguishable in both liquids. 

\clearpage
\section{Orientational Order}
\begin{figure}[h!]
\includegraphics{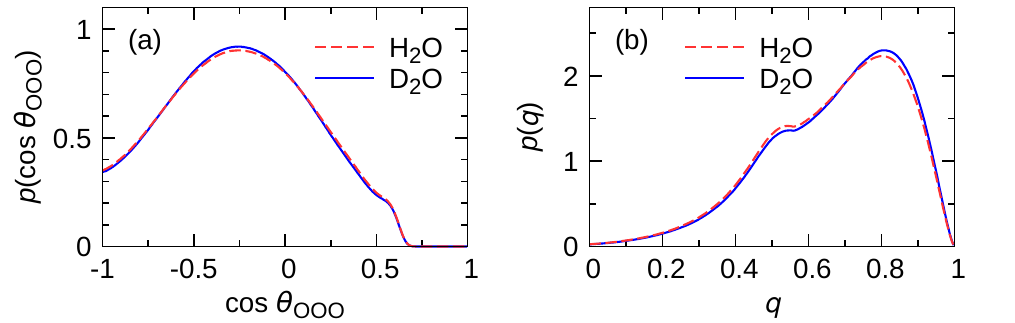}
\caption{ 
Orientational
ordering of H$_2$O compared to D$_2$O. (a) Probability density of the cosine of the oxygen triplet angle ($\theta_{\mathrm{OOO}}$); $\theta_{\mathrm{OOO}}$ is computed for O atoms inside the first solvation shell of a central O atom.
(b) Probability density of the tetrahedral order parameter.
See text for definitions. 
}
\label{structure}
\end{figure}

The orientational ordering of H$_2$O and D$_2$O is probed by the oxygen triplet angle $\theta_{\mathrm{OOO}}$, which is the included angle $\theta_{ijk}$ between the lines connecting O$_j$ to O$_i$ and O$_k$, where O$_i$ and O$_k$ are within a cutoff distance $d$ of O$_j$;
$d$ is chosen as the distance where the average O--O coordination number is equal to 4 following earlier literature \cite{Soper2008Quantum}.
For a perfectly tetrahedral arrangement, $\theta_{\mathrm{OOO}}$ is equal to 
$\approx$~109.5$^{\circ}$.
Additionally, the orientational ordering is quantified using the tetrahedral order parameter \cite{Errington2001Relationship}, defined for oxygen atom $j$ by 
\begin{equation} q_j = 1 - \frac{3}{8} \sum_{i=1}^3 \sum_{k = i+1}^4 \left( \cos \theta_{ijk} + \frac{1}{3} \right) , \end{equation} 
where the sums are over the 4 nearest-neighbor oxygen atoms and $\theta_{ijk}$ is defined as above.
Figure \ref{structure} shows the probability density of these two properties for H$_2$O and D$_2$O. 
D$_2$O displays slightly more peaked distributions than H$_2$O
implying that it is the slightly more structured liquid,
although differences are very small.
While a previous experiment reported the same qualitative difference between light and heavy water for orientational ordering \cite{Soper2008Quantum}, the difference that we find here is smaller.
That experimental study relied on 
the
empirical potential structure refinement procedure,
EPSR,
applied to the experimental scattering data, 
which may introduce some 
bias or restrictions 
into the data analysis procedure 
as vividly discussed in the literature 
\cite{Soper2012Comment, Zeidler2012Reply, Soper2013Radial}. 
It might well be that re-analysis of existing data could 
lead to a smaller isotope effect on the tetrahedral
order parameter $q$, as it did previously 
in case of the maximum of the intermolecular O$\cdots$H~peak
of light relative to heavy water~\cite{Soper2013Radial}
as discussed in the main text.

\clearpage
\section{H-Bonding Coordinates}
\begin{figure}[h!]
\includegraphics{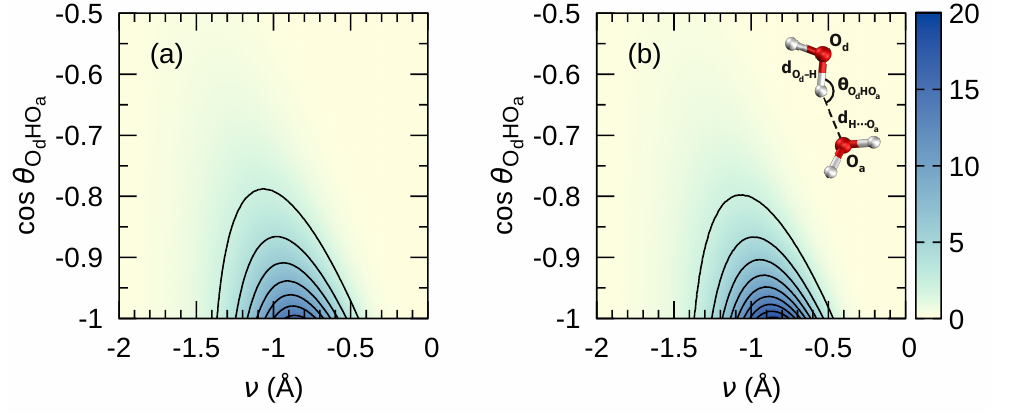}
\caption{
Normalized joint probability densities
of H-bonding coordinates $\nu$ and $\cos \theta_{\mathrm{O_dHO_a}} $ in (a) H$_2$O and (b) D$_2$O.
$\nu = d_{\mathrm{O_d-H}}-d_{\mathrm{H\cdots O_a}}$, and other quantities are defined in the inset.
Contour lines are drawn at intervals of~2.}
\label{proton_transfer_SI}
\end{figure}

The H-bonding coordinates $\cos \theta_{\mathrm{O_dHO_a}}$ and $\nu$, illustrated in Figure \ref{proton_transfer_SI}, were computed 
for each O$_\mathrm{d}$--H$\cdots$O$_{\mathrm{a}}$ triplet, 
where O$_{\mathrm{d}}$ denotes the H-bond donating O atom, and O$_{\mathrm{a}}$ is the H-bond accepting O atom.
The O$_\mathrm{d}$--H$\cdots$O$_{\mathrm{a}}$ triplets are determined by considering in turn every O atom in the system as O$_{\mathrm{d}}$.
For each of the two H (D) atoms covalently bonded to O$_{\mathrm{d}}$,
O$_{\mathrm{a}}$ is selected as the neighboring O atom that minimizes the 
intermolecular H$\cdots$O (D$\cdots$O) distance.

\begin{figure}[h!]
\includegraphics{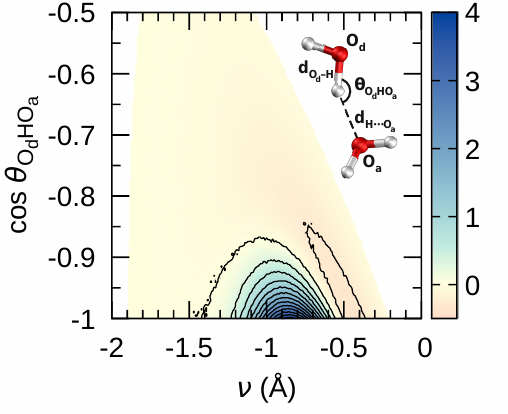}
\caption{
Same figure as Figure 2 in the main text, but with a color scale where the lightness/saturation gradient is similar for positive and negative values.
In Figure 2 in the main text, we have opted for a more steep lightness/saturation gradient for negative values to emphasize 
more clearly
those regions in configuration space where delocalized protons are found.
}
\label{proton_transfer_difference_newscale_SI}
\end{figure}

\clearpage
\section{Self-Diffusion Coefficients}
We computed the self-diffusion coefficients
$D$ of H$_2$O and D$_2$O from RPMD simulations with 64, 96, 128, and 256 molecules using
the molecular centroids~\cite{rpmd-transdiff-2005} via a maximum entropy formalism~\cite{Esser2018Tagging}.
For each of the 
individual RPMD simulations, 
the temperature was determined using the kinetic energy of molecule centroids.
Exploiting the usual Arrhenius-type relationship between temperature and diffusion coefficient,
$D(T) = A \exp [- B/T]$,
the diffusion coefficient at the target temperature of 298~K was
not computed as a simple average of all 
diffusion coefficients
from RPMD initial conditions sampled from the canonical PILE simulations, but extracted
from an exponential fit of $D(T)$ from all 
RPMD runs
as stimulated by~\cite{Munoz-Santiburcio2022Accurate};
the error 
bars are obtained from the scatter of $D(T)$ from the RPMD runs.
The values of $D(298 {\rm\; K})$ for the different cubic box sizes $L$ were used to
explicitly
extrapolate the diffusion coefficient to the infinite box size result
using the $1/L$ relation~\cite{Dunweg1993Molecular, Yeh2004System}
as depicted in Figure~3 of the main text.
We note that we previously~\cite{Daru2022Coupled} applied
the simple semi-empirical extrapolation based on a single system size~$L$,
\begin{equation}
D(L) = D(\infty) - \xi \frac{k_{\rm B} T}{6 \pi \eta L}
\enspace , 
\end{equation}
in conjunction with the experimental shear viscosity $\eta$.
Using 128~H$_2$O molecules, we obtained for light water
\mbox{0.244 {\AA}$^2$} ps$^{-1}$ previously~\cite{Daru2022Coupled},
whereas our improved explicit extrapolation approach provides
\mbox{0.234 {\AA}$^2$ ps$^{-1}$.}

\clearpage
\section{Orientational Relaxation Times}
The orientational relaxation times of the O--H bond vector were obtained from the $C_2^{\mathrm{OH}}(t)$ correlation functions computed for liquid H$_2$O and D$_2$O  with 256 molecules from RPMD simulations following the pioneering work~\cite{Miller2005Quantum}.
As introduced in earlier literature, we use two different approaches
to compute the corresponding $\tau_2$, namely
by fitting an exponential function to $C_2^{\mathrm{OH}}(t)$ in the interval
\mbox{4--15 ps} \cite{Miller2005Quantum, Habershon2009Competing} and by integrating
the correlation function from \mbox{0 to 40 ps} \cite{Wilkins2017Nuclear},
thus following the indicated literature;
our error bar estimates have been computed as described previously~\cite{Daru2022Coupled}.
The orientational relaxation times found through the exponential fit can be compared with results obtained in infrared pump-probe experiments, 
IR, 
and those obtained through 
integration
can be compared to experiments using nuclear magnetic resonance relaxation, 
NMR.
Table~I of the main text compares our CCMD results to the available IR and NMR
experiments.

\clearpage
\section{Orientational Relaxation Times from Experiments}
Here we describe the studies from which we extracted the 
experimental
O--H~bond vector orientational relaxation times $\tau_2^{\mathrm{OH}}$ presented in Table~I in the main text.

Firstly, the reorientation times of H$_2$O and D$_2$O were extracted from NMR~proton and deuteron spin-lattice relaxation experiments~\cite{Jonas1976Molecular}.
We report the reorientation times at 298~K 
obtained
by interpolating the experimental data at 283, 303 and 363~K using an Arrhenius expression~\cite{Bakker2010Vibrational}.
Secondly, using proton NMR~relaxation of H$_2^{17}$O, the 
correlation time 
for reorientation of the O--H~vector 
(in H$_2^{16}$O) was determined to be \mbox{1.71 $\pm$ 0.07 ps} 
at 298 K, and from the $^{17}$O relaxation rate in D$_2^{17}$O, 
the ratio of D$_2$O and H$_2$O correlation times was found to be
\mbox{1.28 $\pm$ 0.02} \cite{Lankhorst1982Determination}.
Assuming isotropic water reorientation, the O--D correlation time is then 
\mbox{2.19 $\pm$ 0.10 ps.} 
Lastly, NMR relaxation data for $^{17}$O in H$_2^{17}$O and D in D$_2$O were obtained at a range of temperatures~\cite{Qvist2012Rotational}. 
The measured rotational correlation times $\tau_{\rm R}$ were fitted to a temperature-dependent power law expression, from which we obtain the correlation times at 298~K.
Note that
$\tau_{\rm R}(^{17}\mathrm{O})$ can be written as the product of $S_{\rm V}^2(^{17}\mathrm{O})$, which is a measure of the amplitude of sub-picosecond librational motions, and $\tau_{\rm s}(^{17}\mathrm{O})$, the structural correlation time \cite{Qvist2012Rotational}.
The structural correlation times from $^{17}$O relaxation, $\tau_{\rm s}(^{17}\mathrm{O})$, and O--H relaxation, $\tau_{\rm s}(\mathrm{O{-}H})$, agree to a good approximation,
so $\tau_{\rm R}(\mathrm{O{-}H}) = \left[ S_{\rm V}^2(\mathrm{O{-}H}) / S_{\rm V}^2(^{17}\mathrm{O}) \right] \tau_{\rm R}(^{17}\mathrm{O})$.
The temperature-dependent $S_{\rm V}^2(\mathrm{O{-}H})$ and $S_{\rm V}^2(^{17}\mathrm{O})$ were obtained from model fits to the correlation times from classical molecular dynamics simulations of SPC/E~water~\cite{Qvist2012Rotational}.
Finally, at 298 K, \mbox{$\tau_{\rm R}(\mathrm{O{-}H}) = 1.83 \pm 0.05$ ps}.
The  D~integral relaxation time $\tau_{\rm R}(\mathrm{D})$ is approximately equal to the O--D integral relaxation time $\tau_{\rm R}(\mathrm{O{-}D})$ \cite{Qvist2012Rotational}, giving that \mbox{$\tau_{\rm R}(\mathrm{O{-}D}) = 2.21 \pm 0.06$ ps} at 298 K.

Complimentary to the NMR~relaxation data are results for the orientational relaxation times from pump-probe IR~experiments.
In these experiments, O--H~reorientation is probed in D$_2$O, and O--D~reorientation in H$_2$O
using isotopically dilute solutions of HDO.
For time scales less than 200~fs, the O--H (or O--D) reorientation is due to librational motion, which does not involve breaking of H-bonds.
In this regime, the motion of the reorienting group is affected by its moment of inertia, and thus we expect \mbox{O--D}~reorientation to be slowed down compared to O--H~reorientation.
For longer time scales, reorientation of O--H/O--D involves rearrangements of the H-bonding network, and therefore the solvent is the most important factor in determining the orientational relaxation times~\cite{Bakker2010Vibrational}.
Thus, to draw conclusions about the orientational relaxation times of H$_2$O, we consider experiments that looked at O--D~reorientation in H$_2$O, 
and for the orientational relaxation times of D$_2$O we consider experiments probing~O--H in D$_2$O
both obtained from the HDO~chromophores.
Using femtosecond mid-infrared pump-probe spectroscopy, the O--D~bond of HDO in H$_2$O was found to have an orientational relaxation time of \mbox{2.5 $\pm$ 0.1 ps} \cite{Rezus2005Orientational}.
In that study, the authors compared their finding to the O--H~orientational relaxation time in D$_2$O from earlier work using IR~pump-probe spectroscopy, which was \mbox{3.0 $\pm$ 0.5 ps} \cite{Cringus2004Hydrogen}.
Work on HDO in H$_2$O and D$_2$O employing a similar methodology found the same reorientation time constants of \mbox{2.5 $\pm$ 0.2 ps} in H$_2$O, and \mbox{3.0 $\pm$ 0.3 ps} in D$_2$O \cite{Bakker2008Molecular}.

\clearpage

\setlength{\bibsep}{0.0cm}
\noindent
\bibliography{isotope_effects_SI}